\documentclass[aps,pra,onecolumn,groupedaddress,showpacs]{revtex4}

\usepackage{amssymb,amsmath}
\usepackage{graphicx}

\newcommand{\om}{\omega}

\newcommand{\pa}{\partial}

\begin{document}

\title{Two-component nonlinear wave of the cubic Boussinesq-type equation}

\author{G. T. Adamashvili}
\affiliation{Technical University of Georgia, Kostava str.77, Tbilisi, 0179, Georgia.\\ email: $adamash77@gmail.com.$ }

\begin{abstract} The general form of the cubic Boussinesq-type equation is considered. In the special cases this equation is reduced to the three different version of the cubic Boussinesq equations and also generalized modified cubic  Boussinesq equation. Using the slowly varying envelope approximation and the  generalized perturbation reduction method the cubic Boussinesq-type equation is transformed to the coupled nonlinear  Schr\"odinger equations and the two-component nonlinear solitary wave solution is obtained. Explicit analytical expressions for the shape and parameters of the two-component nonlinear pulse oscillating with the sum and difference of the frequencies and wave numbers are presented. It is shown that obtained solution coincide with the vector $0\pi$ pulse of the self-induced transparency.

\vskip+0.2cm
\emph{Keywords:} Two-component nonlinear waves, Generalized perturbation reduction method, Cubic Boussinesq-type equation.
\end{abstract}


\maketitle

\section{Introduction}

The main object of research in the physics of nonlinear waves are nonlinear solitary waves of a stationary profile. Nowadays, solitary waves are under intense investigation in many nonlinear systems including optics, acoustics, plasma, hydrodynamics, etc. Usually, two basic types of solitary waves are considered: single-component (scalar) and two-component (vector) nonlinear waves. These waves differ significantly both in their properties and in the mathematical methods of their investigation. Scalar single-component nonlinear waves are single waves, propagating in such a way that their parameters and profile are preserved. Among single-component solitary waves, soliton and breather are considered quite often. The two-component vector pulse is a bound state of two nonlinear wave packets with same velocities and different polarizations (for waveguide mode) or different frequencies and wave numbers with the same or various polarizations[1-11].

Depending on the physical situation and the properties of the medium in which the nonlinear wave propagates, one-component or/and two-component waves can be formed. There are effects that cannot be described using only one-component waves, and for an adequate description of the physical process, the concept of two-component waves is necessary. One of the most bright such effect is self-induced transparency (SIT) [12-14]. On the other hand, not in any physical systems can be form two-component waves, and to determine the possibility of the formation of two-component waves, a special study is required.

Of particular interest are two-component vector breathers (breather molecules). Breather molecule, is bound breather pair with identical polarization. One breather oscillates with the sum and the second with the difference of the frequencies and wave numbers (SDFW). In the theory of SIT, such a nonlinear two-component solitary wave is called the vector $0\pi$ pulse, which was first studied in optics and then in acoustics using the generalized perturbative reduction method proposed in Refs.[14-16]. Later such vector two-component waves also were studied for waves of completely different nature and in various fields of physics such as plasma physics, hydrodynamics, field theory, metamaterials, etc [8-10].

In the theory of nonlinear waves depending from the duration of the pulses we have to consider two different type processes and corresponding the partial nonlinear differential equations which describe them. At this we have to use different mathematical methods of the solution of the nonlinear equations. To the first case belong the relatively wide width pulses we can use the slowly varying envelope approximation [12,17] and we have possibility to consider great class different physical phenomena in the various fields of research. In the second case we can consider very short pulses, for instance, ultra-fast phenomena for the few-cycle pulses in optics [18].

After discover the nonlinear solitary waves - solitons its properties have been described by some well known nonlinear partial differential equations, such as the Korteweg-de Vries equation, the scalar nonlinear Schr\"odinger equation and the Boussinesq equation among others. In the beginning the nonlinear solitary waves have been studied in the shallow water waves but later these waves have been widely investigated in various fields of physics for different nature of waves.

The nonlinear properties of shallow water waves can be modeled by various nonlinear partial differential equations. These include the Benjamin-Bona-Mahony equation, the Korteweg-de Vries equation and different versions of the Boussinesq equations, among others [9, 19-23]. There are various types of solutions that are revealed for these nonlinear equations. These solutions include the nonlinear single-component (scalar) solitary waves. In addition to single-component nonlinear waves, two-component waves, such as vector breathers, are also considered. Special type two-component vector breather oscillating with the sum and difference of the frequencies and wave numbers have been considered  using the generalized perturbative reduction method.

The generalized Boussinesq equation has the form [23, 24]
\begin{equation}\label{eq1}
 \frac{\partial^{2} U}{\partial t^{2}}- C \frac{\partial^{2} U}{\partial z^{2}}  - D  \frac{\partial^{4} U}{\partial z^{4}}+ G   \frac{\partial^{2} U^{n}}{\partial z^{2}} =0,
\end{equation}
where $U(z, t)$ is a real function of space coordinate $z$ and time $t$ and represents the wave profile, while  $C,\;D $ and $G $ are arbitrary constants.

When $n = 3,$  Eq.(1) is reduced to the cubic Boussinesq equation
\begin{equation}\label{e2}
 \frac{\partial^{2} U}{\partial t^{2}}- C \frac{\partial^{2} U}{\partial z^{2}}  - D  \frac{\partial^{4} U}{\partial z^{4}}+ G   \frac{\partial^{2} U^{3}}{\partial z^{2}} =0,
\end{equation}
or in the dimensionless form
\begin{equation}\label{eq2}\nonumber
 \frac{\partial^{2} U}{\partial t^{2}}-  \frac{\partial^{2} U}{\partial z^{2}}  -   \frac{\partial^{4} U}{\partial z^{4}}+ 2  \frac{\partial^{2} U^{3}}{\partial z^{2}} =0.
\end{equation}

Eq.(2) is a well-known model of dispersive nonlinear waves which describes properties of nonlinear solitary waves in a one dimensional lattice and in shallow water under gravity. The Boussinesq equation is also used in the analysis of many other phenomena, for instance, in geotechnical engineering practice [25, 26].
There are several well-known methods which can be used to solve the cubic Boussinesq equation. The tanh method, the variational iteration method, and several other
approaches to solve Eq.(2) and to analyze the solitary waves have been applied [24, 27, 28].

Sometimes, another form of the cubic Boussinesq equation when the term ${\partial^{4} U}/{\partial z^{4}}$ is replaced by the term ${\partial^{4} U}/{{\partial z^{3}}{\partial t}}$  is also considered [29]
\begin{equation}\label{eq3}
 \frac{\partial^{2} U}{\partial t^{2}}- C \frac{\partial^{2} U}{\partial z^{2}}  - D  \frac{\partial^{4} U}{{\partial z^{3}}{\partial t}}+ G   \frac{\partial^{2} U^{3}}{\partial z^{2}} =0.
\end{equation}

The modified improved Boussinesq  equation
\begin{equation}\label{mibe}
 \frac{\partial^{2} U}{\partial t^{2}}-\frac{\partial^{2} U}{\partial z^{2}}- \frac{\partial^{4} U}{{\partial z^{2}}{\partial t^{2}}}-a  \frac{\partial^{2} U^{3}}{\partial z^{2}}=0,
\end{equation}
is one more a modified type of Boussinesq equations for the modelling water-wave problems in weakly dispersive medium such as surface waves in shallow waters or ion acoustic waves [30].

The generalized modified Boussinesq equation is [31]
\begin{equation}\label{MBA}
 \frac{\partial^{2} U}{\partial t^{2}}- \vartheta  \frac{\partial^{4} U}{{\partial t^{2}}{\partial \xi^{2}}}= \frac{\partial^{2} f(U)}{\partial \xi^{2}},
\end{equation}
where $\vartheta$ is a constant and $f(U)$ is an arbitrary nonlinear function, $\xi$ is the the space coordinate. Eq.(5) describes the wave propagation of elastic rods and also the nonlinear lattice modes, iron sound waves, and vibrations in a nonlinear string.

In the system of coordinates moving along  axis $\xi$ with velocity $V$, the equation (5) is transformed to the following equation
\begin{equation}\label{MBA1}
  \frac{\partial^{2} U}{\partial t^{2}}+V^{2}\frac{\partial^{2} U}{\partial z^{2}}-2 V \frac{\partial^{2} U}{{\partial z}{\partial t}} +V^{2 }\vartheta\frac{\partial^{4} U}{\partial z^4} - 2 V\vartheta \frac{\partial^{4} U}{{\partial z^{3}}{\partial t}}+\vartheta \frac{\partial^{4} U}{{\partial z^{2}}{\partial t^{2}}}=\frac{\partial^{2} f(U)}{\partial z^{2}},
\end{equation}
where
$$
\xi=z-V t,\;\;\;t=t.
$$

We consider the following nonlinear cubic Boussinesq-type  equation which  incorporate the above presented different versions of the Boussinesq-type equations (2), (3), (4) and (6):
\begin{equation}\label{gb}
 \alpha \frac{\partial^{2} U}{\partial t^{2}}+\beta\frac{\partial^{2} U}{\partial z^{2}}+ \gamma\frac{\partial^{2} U}{{\partial z}{\partial t}} +\delta\frac{\partial^{4} U}{\partial z^4} +\mu \frac{\partial^{4} U}{{\partial z^{3}}{\partial t}}+\nu \frac{\partial^{4} U}{{\partial z^{2}}{\partial t^{2}}}=- G   \frac{\partial^{2} U^{3}}{\partial z^{2}},
\end{equation}
where we suppose that the nonlinear arbitrary function
\begin{equation}\label{fg}\nonumber
f(U)=-G U^{3}.
\end{equation}

At the consideration of the nonlinear solitary wave solutions of the nonlinear partial differential equations it is very important role play the pulse duration. In the case for the relatively wide pulses with the width of $T$ which satisfies the condition $\omega T >> 1$, where $\omega$ is the carrier wave frequency. For such waves we can use the slowly varying envelope approximation [12, 17] and in this case we can represent the function $U(z, t)$ in the form
\begin{equation}\label{eq4}
U(z,t)=\sum_{l=\pm1}\hat{U}_{l}(z,t) Z_l,\;\;\;\;\;\;\;\;\;\;Z_l=e^{{il(k z -\om t)}}
\end{equation}
where $Z_{l}= e^{il(kz -\om t)}$ is the fast oscillating function, $\hat{U}_{l}$ are the slowly varying complex envelope functions, which satisfied inequalities
\begin{equation}\label{swa}
 \left|\frac{\partial \hat{U}_{l}}{\partial t}\right|\ll\omega
|\hat{U}_{l}|,\;\;\;\left|\frac{\partial \hat{U}_{l}}{\partial z
}\right|\ll k|\hat{U}_{l}|.
\end{equation}
$k$ is the wave number of the carrier wave. For the reality of $U$, we set: $ \hat{U}_{+1}= \hat{U}^{*}_{-1}$.

The purpose of the present work is to consider the two-component solution of the generalized modified cubic Boussinesq-type  equation (7) using generalized perturbative reduction method Eq.(12), when the function $U(z,t)$ satisfies the slowly varying envelope approximation, Eq.(9).

The rest of this paper is organized as follows: Section II is devoted to the cubic Boussinesq-type equation for slowly varying complex envelope function $\hat{U}_{l}$ and  using the generalized perturbation reduction method, we will transform Eq.(7) to the coupled nonlinear  Schr\"odinger equations for auxiliary functions.
In Section III, will be presented the explicit analytical expressions for the shape and parameters of the two-component nonlinear pulse. Finally, in Section IV, we will discuss the obtained results.

\vskip+0.5cm

\section{The generalized perturbative reduction method}

We consider a pulse whose duration satisfies  the condition $T >>\omega^{-1}$.  Substituting  Eq.(8) into (7) we obtain the connection between the parameters $\omega$ and $k$ in the form
\begin{equation}\label{dis}
\alpha {\omega}^{2} + \beta k^{2}-  \gamma k \omega  - \delta k^{4}  + \mu   k^{3} \omega -\nu  \omega^{2} k^{2}  =0
\end{equation}
and the cubic Boussinesq-type equation for envelope function $\hat{U}_{l}$:
\begin{equation}\label{lin2}
\sum_{l=\pm1}Z_l [ il A_{1} \frac{\partial \hat{U}_{l}}{\partial t} + A_{6}\frac{\partial^{2} \hat{U}_{l}}{\partial t^2}-il A_{2} \frac{\pa \hat{U}_{l}}{\pa z}
+A_{3} \frac{\pa^{2} \hat{U}_{l}}{\pa z^2} +A_{5} \frac{\pa^{2} \hat{U}_{l}}{{\pa z}{\partial t}}
+il A_{4}  \frac{\partial^{3} \hat{U}_{l}}{\partial z^{3}} + \delta \frac{\partial^{4} \hat{U}_{l}}{\partial z^{4}} +i l A_{7} \frac{\pa^{3} \hat{U}_{l}}{{\pa z^2}{\partial t }}
$$$$
+ \mu \frac{\partial^{4} \hat{U}_{l}}{{\partial z^{3}}{\partial t}}    +  i l A_{9} \frac{\pa^{3} \hat{U}_{l}}{{\pa z}{\partial t^2}}
 +\nu  \frac{\pa^{4} \hat{U}_{l}}{{\pa z^2}{\partial t^2}}]=F_{non},
\end{equation}
where
$$
 A_{1} =-2 \alpha \omega + \gamma  k  - \mu   k^{3} +2 \nu  \omega k^2,
$$
$$
A_{2}= - 2 \beta  k   + \gamma  \omega  +4 \delta  k^{3} -3 \mu   k^{2} \omega   + 2 \nu     k \omega^2,
$$$$
A_{3}=\beta -6 \delta  k^{2} + 3 \mu  k \omega    - \nu  \omega^2,
$$
$$
A_{4} =   4 \delta  k -\mu  \omega,
$$
$$
A_{5}= \gamma  -3 \mu  k^{2}   +  4 \nu   k \omega,
$$
$$
A_{6}=\alpha -k^2 \nu,
$$
$$
 A_{7}= 3 \mu k   -2 \nu  \omega,
$$
$$
 A_{9}=   2 \nu   k.
$$
The nonlinear term of Eq.(11) equal to
$$
F_{non}=-3 G  \sum_{l=\pm1} \frac{\partial^{2} }{\partial z^{2}}(Z_l \hat{U}_{l}^{2}\hat{U}_{-l}).
$$

In order to consider the two-component vector breather solution of the Eq.(2), we use the generalized perturbative reduction method
(see, for instance [8, 9] and references therein) which makes it possible to transform the cubic Boussinesq-type equation for the functions $\hat{U}_{l}$ to the coupled nonlinear Schr\"odinger  equations for auxiliary functions $f_{l,n}^ {(\alpha)}$.
As a result, we obtain a two-component nonlinear pulse oscillating with the  difference and sum of the frequencies and wave numbers. In the frame of this method, the complex envelope function  $\hat{U}_{l}$ can be represented as
\begin{equation}\label{gprm}
\hat{U}_{l}(z,t)=\sum_{\alpha=1}^{\infty}\sum_{n=-\infty}^{+\infty}\varepsilon^\alpha
Y_{l,n} f_{l,n}^ {(\alpha)}(\zeta_{l,n},\tau),
\end{equation}
where $\varepsilon$ is a small parameter,
$$
Y_{l,n}=e^{in(Q_{l,n}z-\Omega_{l,n}
t)},\;\;\;\zeta_{l,n}=\varepsilon Q_{l,n}(z-v_{{g;}_{l,n}} t),
$$$$
\tau=\varepsilon^2 t,\;\;\;
v_{{g;}_{l,n}}=\frac{\partial \Omega_{l,n}}{\partial Q_{l,n}}.
$$

It is assumed that the quantities $\Omega_{l,n}$, $Q_{l,n}$ and $f_{l,n}^{(\alpha)}$ satisfies the inequalities for any $l$ and $n$:
\begin{equation}\label{rtyp}\nonumber\\
\omega\gg \Omega_{l,n},\;\;k\gg Q_{l,n},\;\;\;
\end{equation}
$$
\left|\frac{\partial
f_{l,n}^{(\alpha )}}{
\partial t}\right|\ll \Omega_{l,n} \left|f_{l,n}^{(\alpha)}\right|,\;\;\left|\frac{\partial
f_{l,n}^{(\alpha )}}{\partial \eta }\right|\ll Q_{l,n} \left|f_{l,n}^{(\alpha )}\right|.
$$

Substituting Eq.(12) into Eq.(11), for the left-hand side of the cubic Boussinesq-type equation we obtain
\begin{equation}\label{eqz}
\sum_{l=\pm1}\sum_{\alpha=1}^{\infty}\sum_{n=\pm 1}\varepsilon^\alpha Z_{l} Y_{l,n}[W_{l,n}
+ \varepsilon i J_{l,n}\frac{\partial }{\partial \zeta_{l,n} } - \varepsilon^2 i l \hat{h}_{l,n}  \frac{\partial }{\partial \tau}
-\varepsilon^{2} Q^{2}_{l,n} H_{l,n}\frac{\partial^{2} }{\partial \zeta_{l,n}^{2}}+O(\varepsilon^{3})]f_{l,n}^{(\alpha)}=F_{non},
\end{equation}
where
\begin{equation}\label{cof}
W_{l,n}= l A_{1} n\Omega -  A_{6} {\Omega}^{2}+ l A_{2} n Q -A_{3}  Q^{2} +A_{5}  Q \Omega+ l A_{4} n Q^{3} + \delta  Q^{4} - l A_{7} n Q^{2} \Omega  - \mu Q^{3} \Omega+ l A_{9} n {\Omega}^{2}Q    +\nu   Q^{2} \Omega^{2},
$$
$$
J_{l,n}= - l A_{1} Q v_g  +2 A_{6}   n \Omega Q v_g  - l A_{2}  Q + 2 A_{3}   n Q^{2} - A_{5}   n Q (Q v_g +\Omega)  -l A_{4}  3  Q^{3}  - 4 \delta  n Q^{4}  +  l A_{7}  Q^{2}(Q v_g +  2 \Omega)
$$$$
+  \mu n Q^{3} ( Q v_g +3 \Omega)-  l A_{9}    Q \Omega ({\Omega} + 2    Q v_g ) -2\nu    n Q^{2} \Omega( Q v_g + \Omega),
$$
$$
\hat{h}_{l,n}=l n ( 2  A_{6} \Omega - A_{5} Q   + \mu Q^{3}   -  2\nu  Q^{2} \Omega) + A_{7}   Q^{2}-2 A_{9} Q \Omega -  A_{1},
$$
$$
H_{l,n}=- A_{6}  v_g^{2}- A_{3} +  \nu  Q^{2} v^{2}_g    +  4 \nu \Omega  Q v_g  + \nu \Omega^{2}    + A_{5} v_g    + 6 \delta Q^{2}   -3\mu  Q ( Q v_g + \Omega)
$$$$
+  l n [   2  \Omega A_{9}    v_g  + Q  A_{9} {v}^{2}_{g}    +  3  A_{4}   Q -   2 Q A_{7}  v_g  - A_{7} \Omega].
\end{equation}

For the sake of simplicity, we omit $l$ and $n$ indexes for the quantities $\Omega_{l,n}$, $Q_{l,n}$, ${v_{g;}}_{l,n}$ and $\zeta_{l,n}$ in Eqs.(14) and furthermore where this will not be messy.

The nonlinear term $F_{non}$ on the right-hand side of the equation (13) is of order to $\varepsilon^{3}$.

To following to the standard procedure for the assimptotic methods we equating to zero, the terms with the same powers of $\varepsilon$. From Eq.(13) we obtain a series of equations. In the first order of $\varepsilon$, we have that when $ f_{l,n}^{(1)}\neq0, $ the connection between of the parameters $\Omega_{l,n}$ and $Q_{l,n}$ has the form
\begin{equation}\label{diss}
 \delta  Q^{4} -  A_{6} {\Omega}^{2} - A_{3}  Q^{2} + A_{5}  Q \Omega  - \mu Q^{3} \Omega      +\nu   Q^{2} \Omega^{2}+ ln( A_{1} \Omega  +  A_{2}  Q  +  A_{4}  Q^{3}  -  A_{7}  Q^{2} \Omega +  A_{9}  {\Omega}^{2}Q) =0.
\end{equation}

From Eq.(15) we obtain
\begin{equation}\label{vg}
v_{{g;}_{l,n}}=\frac{  l n ( A_{2} + 3  A_{4}  Q^{2} - 2  A_{7}  Q \Omega +  A_{9}  {\Omega}^{2} )  - 2 A_{3}  Q + A_{5}   \Omega  + 4\delta  Q^{3}  - 3 \mu Q^{2} \Omega  +2 \nu   Q \Omega^{2}    }{   l n ( A_{7} Q^{2} - A_{1} - 2  A_{9}  {\Omega}  Q) + 2 A_{6} {\Omega} - A_{5}  Q   + \mu Q^{3}  - 2 \nu   Q^{2} \Omega  }.
\end{equation}

From Eqs.(14), (15) and (16), in  the second order of $\varepsilon$, we obtain the equation $J_{l,n}=0$ for any indexes $l$ and $n$.

In the third order of $\varepsilon$,  the cubic Boussinesq-type equation (13), is given by
\begin{equation}\label{l}
\sum_{l=\pm1}\sum_{n=\pm 1}\varepsilon^{3} Z_{l} Y_{l,n}[ - i l \hat{h}_{l,n}  \frac{\partial }{\partial \tau}- Q_{l,n}^{2} H_{l,n}\frac{\partial^{2} }{\partial \zeta_{l,n}^{2}}]f_{l,n}^{(1)}=F_{non}.
\end{equation}

Next we consider the nonlinear term $F_{non}$ proportional to $Z_{+1}$ of the  cubic Boussinesq-type equation which is has to the following  form
\begin{equation}\label{non}
- 3 \varepsilon^{3} G[(k+Q_{+})^{2} ( | f_{+1,+1}^ {(1)}|^{2} +2 | f_{+1,-1}^ {(1)}|^{2} ) Y_{+1,+1} f_{+1,+1}^ {(1)}
$$$$
  +(k-Q_{-})^{2}   ( | f_{+1,-1}^ {(1)}|^{2}   +2   | f_{+1,+1}^ {(1)}|^{2} )Y_{+1,-1} f_{+1,-1}^ {(1)}].
\end{equation}

Combining Eqs.(17) and (18) we obtain the system of nonlinear equations
\begin{equation}\label{2eq}
  i \frac{\partial f_{+1,+1}^{(1)}}{\partial \tau} + Q_{+}^{2} \frac{H_{+1,+1} }{\hat{h}_{+1,+1}} \frac{\partial^2 f_{+1,+1}^{(1)}}{\partial \zeta_{+1,+1} ^2}+\frac{3 G (k+Q_{+})^{2}}{ \hat{h}_{+1,+1}}  ( | f_{+1,+1}^ {(1)}|^{2} + 2 | f_{+1,-1}^ {(1)}|^{2} ) f_{+1,+1}^ {(1)}=0,
$$$$
   i \frac{\partial f_{+1,-1 }^{(1)}}{\partial \tau} + Q_{-}^{2} \frac{H_{+1,-1} }{\hat{h}_{+1,-1}} \frac{\partial^2 f_{+1,-1 }^{(1)}}{\partial \zeta_{+1,-1}^2}+\frac{3 G (k-Q_{-})^{2}}{\hat{h}_{+1,-1}} ( |f_{+1,-1}^ {(1)}|^{2} +2 |f_{+1,+1} ^ {(1)}|^{2} )  f_{+1,-1}^ {(1)}=0.
 \end{equation}
where
\begin{equation}\label{qom}
 Q_{+}=Q_{+1,+1}= Q_{-1,-1},\;\;\;\;\;\;\;\;\;\;\;\;\;\;\;\;\;\;\;\;\;\;\;\;\;  Q_{-}=Q_{+1,-1}= Q_{-1,+1}.
 \end{equation}

\vskip+0.5cm
\section{The two-component vector breather }

After transformation back to the variables $z$ and $t$, from Eqs.(19) we obtain the coupled nonlinear Schr\"odinger equations for the auxiliary functions $\Lambda_{\pm}=\varepsilon  f_{+1,\pm1}^{(1)}$ in the following form
\begin{equation}\label{pp2}
i (\frac{\partial \Lambda_{\pm}}{\partial t}+v_{\pm} \frac{\partial  \Lambda_{\pm}} {\partial z}) + p_{\pm} \frac{\partial^{2} \Lambda_{\pm} }{\partial z^{2}}
+q_{\pm}(|\Lambda_{\pm}|^{2}+2 |\Lambda_{\mp}|^{2} )\Lambda_{\pm}=0,
\end{equation}
where
\begin{equation}\label{pp4}
\Lambda_{\pm}=\varepsilon  f_{+1,\pm1}^{(1)},
$$
$$
p_{\pm}=\frac{H_{+1,\pm 1} }{\hat{h}_{+1,\pm 1}},
$$
$$
 q_{\pm}=\frac{3 G  (k\pm Q_{\pm })^{2}}{ \hat{h}_{+1,\pm 1}},
$$
$$
v_{\pm }= v_{g;_{+1,\pm 1}}.
\end{equation}

The solution of Eq.(21) is given by [8, 11, 16]
\begin{equation}\label{ue1}
\Lambda_{\pm }=\frac{A_{\pm }}{T}Sech(\frac{t-\frac{z}{V_{0}}}{T}) e^{i(k_{\pm } z - \omega_{\pm } t )},
\end{equation}
where $A_{\pm },\; k_{\pm }$ and $\omega_{\pm }$ are the real constants, $V_{0}$ is the velocity of the nonlinear wave. We assume that
$k_{\pm }=\frac{V_{0}-v_{\pm}}{2p_{\pm}}<<Q_{\pm }$  and $\omega_{\pm }<<\Omega_{\pm }.$

Combining Eqs. (8), (12) and (23), we obtain the two-component vector breather solution of the cubic Boussinesq-type equation (7) in the following form:
\begin{equation}\label{vb}
U(z,t)=R \;Sech(\frac{t-\frac{z}{V_{0}}}{T})\{   \cos[(k+Q_{+}+k_{+})z
-(\omega +\Omega_{+}+\omega_{+}) t]
$$$$
+\sqrt{\frac{p_{-}q_{+}-2 p_{+}q_{-}}{p_{+}q_{-}- 2p_{-}q_{+}}}  \cos[(k-Q_{-}+k_{-})z -(\omega -\Omega_{-}+\omega_{-})t]\},
\end{equation}
where
$R=\frac{2 A_{+}}{T}$ is amplitude of the pulse.  Connection between the width $T$ and the velocity $V_{0}$  of the two-component nonlinear pulse is determined as:
\begin{equation}\label{rrw}
T^{-2}=V_{0}^{2}\frac{v_{+}k_{+}+k_{+}^{2}p_{+}-\omega_{+}}{p_{+}},
\end{equation}
where the connection between parameters of the nonlinear wave have the form
\begin{equation}\label{ttw}
A_{+}^{2}=\frac{p_{+}q_{-}- 2p_{-}q_{+}}{p_{-}q_{+}-2 p_{+}q_{-}}A_{-}^{2},
$$$$
 \Omega_{+}=\Omega_{+1,+1}= \Omega_{-1,-1},\;\;\;\;\;\;\;\;\;\;\;\;\;\;\;\;\;\;\;\;\;\;\;\;\;  \Omega_{-}= \Omega_{+1,-1}= \Omega_{-1,+1},
$$$$
\omega_{+}=\frac{p_{+}}{p_{-}}\omega_{-}+\frac{V^{2}_{0}(p_{-}^{2}-p_{+}^{2})+v_{-}^{2}p_{+}^{2}-v_{+}^{2}p_{-}^{2} }{4p_{+}p_{-}^{2}}.
\end{equation}

The connections between parameters $A_{\pm },\; k_{\pm },\;\omega_{\pm }$ and $T$  are given in Refs.\cite{Adamashvili:CPB:21, Adamashvili:Eur.Phys.J.D.:20}.

\vskip+0.5cm

\section{Conclusion}

In this paper, we study the two-component vector breather of the cubic Boussinesq-type equation (7) under the condition of the slowly varying envelope approximation Eq.(9).  We consider nonlinear pulse with the width $T>>\Omega_{\pm }^{-1}>>\omega^{-1}$.
 Using the generalized perturbation reduction method Eq.(12), the Eq.(7) is  transformed to the coupled nonlinear Schr\"odinger equations (21) for the auxiliary functions $\Lambda_{\pm 1}$. As a result, the two-component nonlinear pulse oscillating with the sum and difference  of the frequencies and wave numbers Eq.(24) is formed. The dispersion relation and the connection between parameters $\Omega_{\pm}$ and $Q_{\pm}$ are determined from Eqs.(10) and (15).
 The parameters of the nonlinear pulse from Eqs. (16), (22), (25) and (26) are determined. This solution coincide with the vector $0\pi$ pulse of the self-induced transparency[8].

The cubic Boussinesq-type equation (7) is general equation which united several special cases of the cubic Boussinesq equations.

Indeed, when in Eq.(7) the coefficients satisfied the conditions
\begin{equation}\label{c1}
\alpha=1,\;\;\beta=-C,\;\; \delta=-D,\;\;\gamma=\mu =\nu =0,
\end{equation}
than this equation is transformed to the Eq.(2).

In case when the coefficients in Eq.(7) satisfied the conditions
\begin{equation}\label{c2}
\alpha=1,\;\;\beta=-C,\;\; \mu=-D,\;\;\gamma=\delta =\nu =0,
\end{equation}
than Eq. (7) is coincide with  Eq.(3).

Under the condition when
\begin{equation}\label{c3}
\alpha=1,\;\;\beta=-1,\;\;\gamma=\delta=\mu=0,\;\;\nu=-1,\;\; a=-G,
\end{equation}
than Eq. (7) is reduced to  the Eq.(4).

If the coefficients in Eq.(7) satisfies  the conditions
\begin{equation}\label{c4}
 \alpha=1,\;\;\beta=V^{2},\;\;\gamma=-2V,\;\;\delta=V^{2} \vartheta,\;\;\mu=-2 V \vartheta,\;\;\nu=\vartheta,\;\;f=-G U^{3},
 \end{equation}
in this case Eqs.(6) and (7) are coincides.

The two-component vector breather Eq.(24) we met in the different areas of physics and various nature of waves (see, for instance [8, 9, 15] and references therein). The two-component nonlinear solution Eq. (24) of the cubic Boussinesq-type equation together with scalar one-component solitons of this equation become the theory of the Boussinesq-type nonlinear waves more complete and more widely used.

\vskip+0.5cm

\end{document}